\documentclass[10pt,twocolumn]{article} 
\usepackage{palatino,graphicx}  
\usepackage{psboxit}  

\usepackage[dvipdf]{epsfig} 
\usepackage{epsf}

\begin{document}
\begin{titlepage}
%
\def\nummer{\sf TUD-FI07-01 Februar 2007}
\def\autor{\sf Katrin Borcea-Pfitzmann,\\Anne-Katrin Stange}
\def\institut{\sf Institut f\"ur Systemarchitektur}
\def\titel{Privacy -- an Issue for eLearning? \\A Trend Analysis Reflecting the Attitude of European eLearning Users}

\begin{picture}(160,240)%
%
\put(60,70){\makebox(96,5)[]%
{\parbox{85\unitlength}%
{\begin{center}\sffamily\bfseries \titel \end{center}}}}
%
\put(60,96){\makebox(96,5)[]%
{\parbox{85\unitlength}%
{\begin{center}\sffamily\large\bfseries\em \autor \end{center}}}}
%
\put(60,84){\makebox(96,5)[]%
{\parbox{85\unitlength}%
{\begin{center}\sffamily\small\bfseries \institut \end{center}}}}
%
\put(60,106){\makebox(96,5)[]{\sffamily\normalsize\bfseries \nummer }}
%
%
%
%
\put(5,95){\makebox(50,6)[r]%
{\sffamily\Large\mdseries Technische Berichte}}
%
\put(5,87){\makebox(50,6)[r]%
{\sffamily\Large\mdseries Technical Reports}}
%
\put(5,80){\makebox(50,6)[r]%
{\sffamily\normalsize\mdseries ISSN 1430-211X}}
%
%
%
%
%
%
%
\put(-18,245){\rotatebox{0}{\includegraphics[scale=0.5]{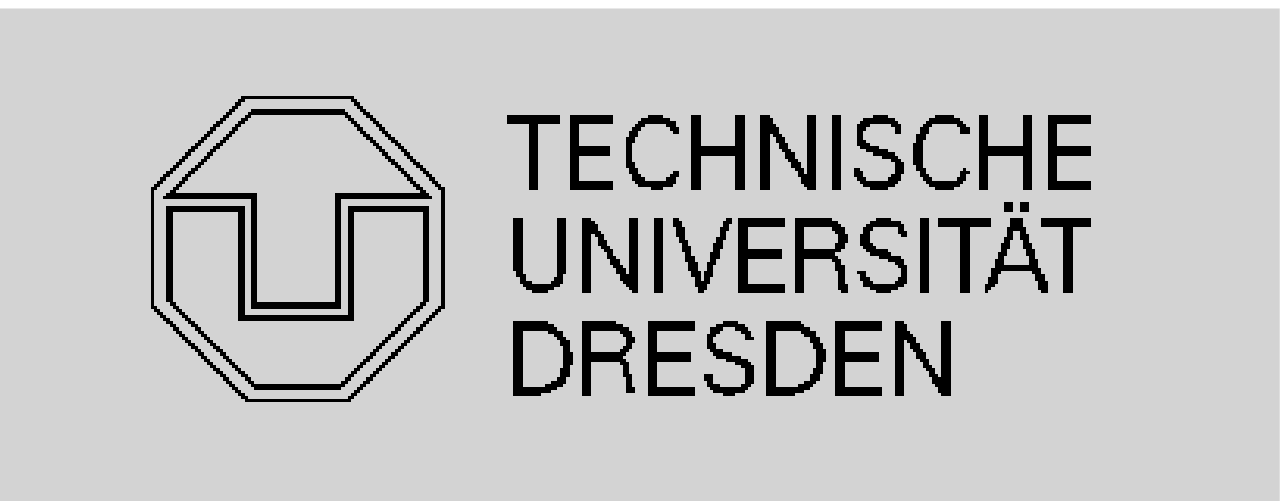}}}%
\put(44,180){\rotatebox{0}{\includegraphics[scale=0.75]{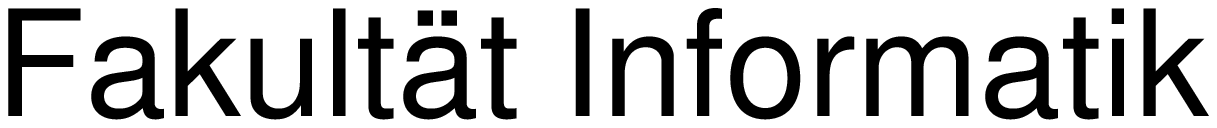}}}%
\put(58,63){\rotatebox{0}{\includegraphics[scale=0.98]{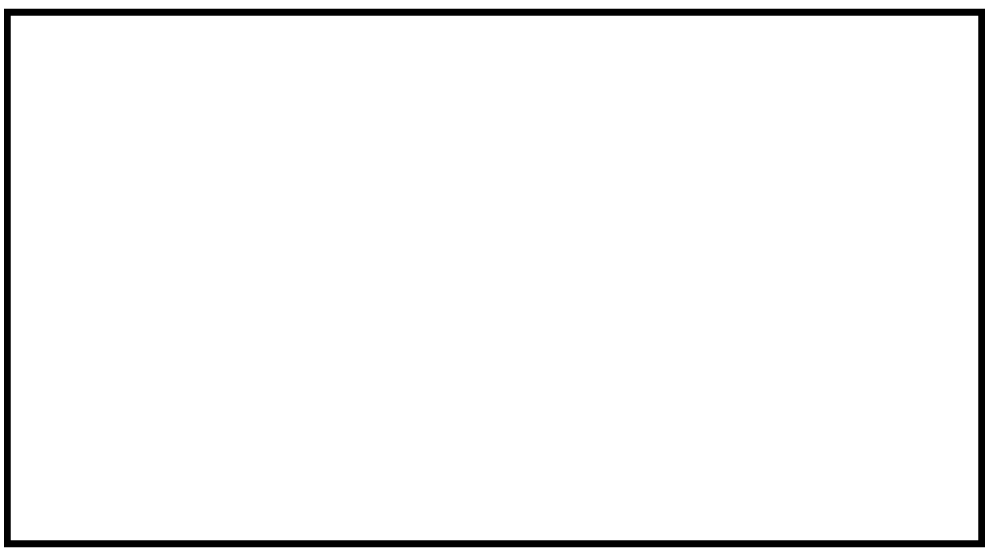}}}%
\put(50,-20){\rotatebox{0}{\includegraphics[scale=0.6]{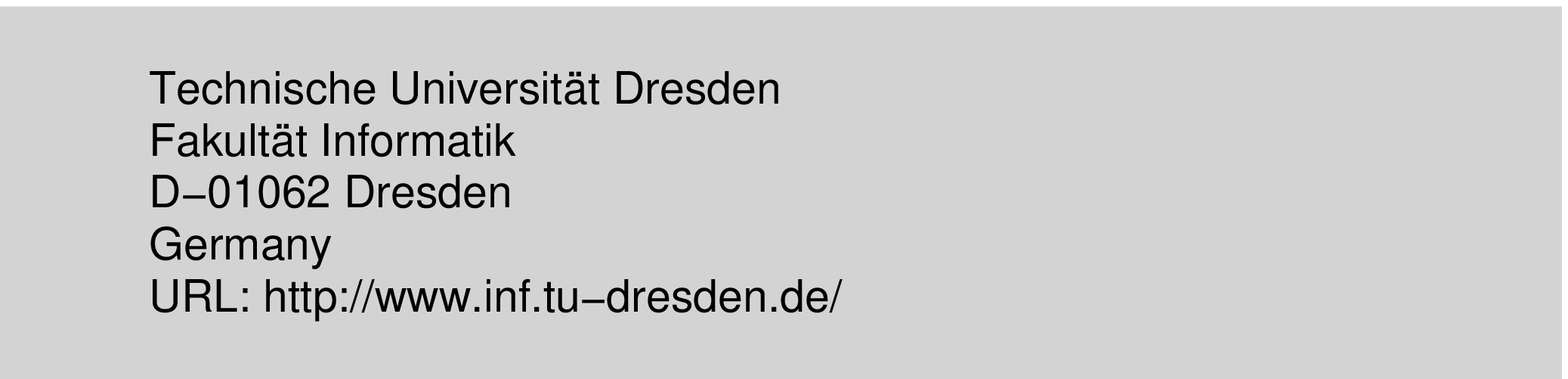}}}%
\put(50,13){\rotatebox{0}{\includegraphics[scale=1.3]{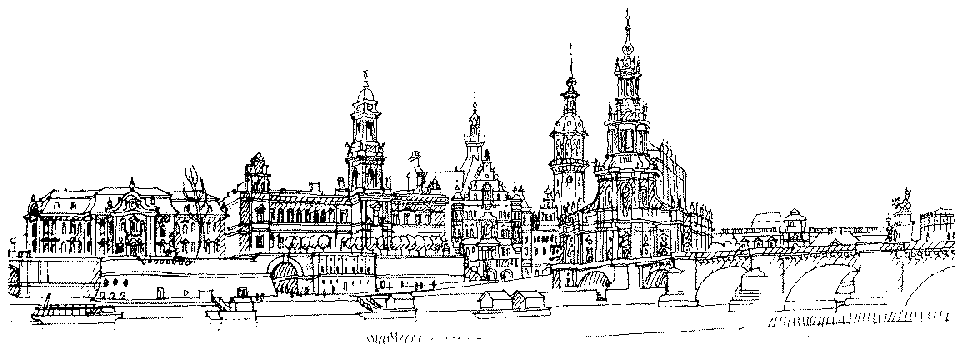}}}
\end{picture}%
\end{titlepage}

\title{Privacy - an Issue for eLearning? A Trend Analysis Reflecting the Attitude of European eLearning Users}

\author{Katrin Borcea-Pfitzmann \hspace{1cm} Anne-Katrin Stange\\
Dresden University of Technology\\ Department of Computer Science\\ Dresden, Germany\\ 
katrin.borcea@tu-dresden.de\hspace{1cm} as024121@inf.tu-dresden.de\\
}
\date{}
\maketitle
\thispagestyle{empty}

\begin{abstract}
Availing services provided via the Internet became a widely accepted means in organising one's life. Beside others, eLearning goes with this trend as well. 
But, while employing Internet service makes life more convenient, at the same time, it raises risks with respect to the protection of the users' privacy. This paper analyses the attitudes of eLearning users towards their privacy by, initially, pointing out terminology and legal issues connected with privacy. Further, the concept and implementation as well as a result analysis of a conducted study is presented, which explores the problem area from different perspectives. The paper will show that eLearning users indeed care for the protection of their personal information when using eLearning services. However, their attitudes and behaviour slightly differ. In conclusion, we provide first approaches of assisting possibilities for users how to resolve the difference of requirements and their actual activities with respect to privacy protection.
\end{abstract}


\section{Introduction: Why Should We Answer the Title Question}
\label{sec:IntroductionWhyDoWeEngageOurselvesInAnsweringTheTitleQuestion}

When watching at the developments of information technology, one can state that the previous years were characterised by a rapid upturn of the new and advanced possibilities in the field of communication. However, being busy with getting known the nice features of the new technologies we forgot to care about secondary consequences. These imply the risks in terms of privacy which are very much connected to the new freedom that the communication means deliver. But, not only the technical options for communication have changed. This development had also quite an influence on our social backgrounds as well as on our culture of living. Especially, the last mentioned aspects are important to consider when designing next generation eLearning platforms the characteristics of which will be stronger oriented on collaborative, informal and community-oriented learning scenarios. 

This paper analyses the current attitude of eLearning users towards their privacy when acting on the Internet and, above all, when working with eLearning systems. Thereby, we understand eLearning as comprising all learning applications, which are reachable and usable via the Internet. Those learning applications may be designed for cooperative, collaborative, or self-determined learning.
Especially, in multi-user eLearning environments we have to distinguish relevant tasks like learning, teaching and organising, which are represented by different roles: the learner, the tutor, the mentor, the author, and the administrator.
The interplay of the mentioned roles requires an exchange of personal information, which might be intended but also unintended since each individual has own feelings with respect to protecting his/her privacy or disclosing specific information, respectively.

Given this context, we formulated questions such as: How is privacy being perceived in eLearning situations? Do we have to consider privacy issues when designing eLearning systems, and if so - which way has it to be done?

First, we will start the discussion by reflecting definitions of privacy as well as contrasting this term to related terminology. In order to further motivate the discussion on this topic, we will refer to legal regulations, which exist to protect the users' privacy in the informational society. After that, we will analyse our survey and its results in detail as well as setting them in contrast to related studies. Following, we will conclude the discussion by presenting according consequences for the design of collaborative eLearning environments.

The aim of this paper is to approach the wishes and requirements of eLearning users with respect to their privacy preferences. Therefore, eLearning users from all over Europe got the possibility to participate in the survey, which addressed the above-indicated issues. 

\section{Terminology and Motivation}
\label{sec:TerminologyAndMotivation}

\subsection{Privacy and Informational Self-Determination}
\label{sec:PrivacyAndInformationalSelfDetermination}

People often do not associate the same meaning with the term privacy. That is why we attend to a discussion about the term's proximity that further relates to informational self-determination as well as to security.

In the waning years of the 19th century and beginning of the 20th century, for the first time, privacy became an issue of legal considerations. So, the judges S.D. Warren and L.D. Brandeis have given privacy a legal base by defining it as "the right to be left alone" \cite{rossler.165}. In 1967, A.F. Westin for the first time brought informational privacy into focus by reflecting his perception of privacy as "the claim of individuals, groups or institutions to determine for themselves when, how and to what extend information about them is communicated to others", which means that there should be the possibility to control the access to private information. Thus, in a result of increased data processing and the possibility to store more and more information, informational privacy had encountered the technological progress and the associated risks.

Beate Rößler [Roes01] determines a spectrum of three privacy classes \cite{rossler.165}:
\begin{itemize}
	\item \textit{Local privacy} means the protection of rooms and areas.
	\item \textit{Privacy of actions and behaviour} implies a free choice of, e.g., confession and clothing, but also protection against heteronomy.
	\item \textit{Informational privacy} focuses the protection of information about oneself, opinions about others, and communications with other persons.
\end{itemize}
Below, we especially refer to the class \textit{informational privacy} when addressing \textit{privacy}.

Often, the term \textit{privacy} is used in association with \textit{security}. However, while security means the protection of \textit{all} confidential data and information, privacy comprises the level of desired security of a person's own data and information. We explicitly state "level of desired security" since acting in the informational society is always a trade-off between privacy and the provided opportunities. Thus, e.g., to get a bank credit or medical care one has to disclose personal data in order to be able to participate in the social life.

\subsection{Protection of Privacy on the Internet and within eLearning}
\label{sec:ProtectionOfPrivacyOnTheInternetAndWithinELearning}

When turning to practice, a series of problems come into sight, which is connected with the protection of privacy on the Internet: First, users of online services are often not able to recognise that data about themselves is collected. Often, they have insufficient knowledge about the data processing procedures and, thus, about the concrete purpose of information gathering done by service providers. Such procedures allow for associating the given data with databases of different companies to create a global profile or a pattern of behaviour of that person. For example, cookies, which are used to store session data, can collect information over a long period. Although cookies are stored on the person's computer, this information can be requested from the initiator every time the user connects to the according service. On that basis, it becomes easy for the service provider to analyse the behaviour of the user on the Internet. Second, the "Internet does not forget", i.e. once information had been digitised and stored on a networked storage it becomes hard - if not even impossible - to get it really deleted. 

In order to solve those problems, different possibilities help to control the disclosure of personal information to service providers and other interested institutions: One of the possible countermeasures represents the use of anonymisers, which conceal the IP address of the user's computer and hide the type of transmitted data, e.g. JAP \footnote{http://jap.inf.tu-dresden.de}  or Tor \footnote{http://tor.eff.org/index.html.en}. Another approach consists in special software (such as CookieCooker \footnote{http://www.cookiecooker.de/}) that exchanges the user's cookies with those of other users. Further, technologies of privacy-enhancing identity management, like the ones currently being developed in PRIME \footnote{https://www.prime-project.eu/}, assist the users in managing their partial identities  when using different web services \cite{pfitzmann.117}. 

While analysing the field of privacy in eLearning we encountered that there is a special situation: In order to being able to optimally support the learning processes it is inevitable to indicate information that is personal to the users. This is especially true when collaboration is one of the learning environment's paradigms for knowledge building and exchange. In such a case, group awareness is one of the most important aspects in order to work efficiently together with other participants \cite{kettel.94}. Depending on who the interaction partner is, the users are willing to disclose their real or their nicknames, information about relations to other participants, and their level of performance (grades in tests, certificates). Besides collaboration, personalisation and adaptation of learning environments require to indicate privacy-relevant information as well.

In order to sum up the problem analysis, we need to mention that quite a lot of different personal data plays an important role within eLearning environments depending on the actual context and the interaction partners. Nevertheless, disclosing all of the personal data to everyone might advance risks connected to biased environments, profiling etc. 

\subsection{Approaches of Legal Regulations}
\label{sec:ApproachesOfLegalRegulations}

In the frame of legal regulations, different approaches exist, which are outlined in the following. First, we would like to mention the guidelines of OECD, the Organization for Economic Co-operation and Development, compiled in 1980 \cite{anon.164}. These guidelines represent recommendations of how to protect sensitive data (e.g. by limiting data collections, indicating the purpose of data collection, and by limitation of use). A further aim was also to guarantee secure data traffic to other states than the information origin. The OECD guidelines formed the fundament for the Data Protection Directive of the European Parliament and of the Council (1981). Its objective was to regulate the handling of personal information for almost all European countries based on determined principles with respect to confidentiality, security, and transparency. The unification of all data protection laws of the countries of the European Union is aimed by the implementation of the Convention for the Protection of Individuals with regard to Automatic Processing of Personal Data of the Council of Europe (1995) \cite{anon.160}. Further, we would like to point out an event in Germany called "Volkszählungsurteil" \cite{anon.1}: In 1983, a population census should take place by collecting a large number of personal data (such as telephone number or place of work etc.). Such kind of information collection intended by the "Volkszählungsurteil" \cite{anon.1} was judged as violation of the human rights. Therefore, for the first time, the right for informational self-determination became an accepted claim in a European country.  

In 2003, a representative survey addressing the attitudes of the European citizens \cite{soufflotdemagny.159} reflected that 61\% of the respondents did not hear about the existence and objectives of data protection laws of the EU  while only 31\% responded that they know about the laws. 70\% of the respondents expressed their opinion that people's awareness about personal data protection in their country is rather low. Nevertheless, almost half of the respondents perceived a high level of personal data protection provided by the law in their country. This means, that, in 2003, about two-thirds of the European citizens did not know which rights and possibilities with respect to protecting their privacy they hold. However, they feel that there is a high level of data protection in their countries without knowing what concrete effects this implies. 

Notwithstanding of such regulations, privacy rights are going to get restricted more and more: Providers of telecommunication and information services are obligated to collect and store data about connections and location for a wide range of kinds of electronic data. Thus, caring for one's privacy and informational self-determination is not just an issue of legal regulations but also a personal concern of every user him/herself.

\subsection{Conclusions from the Analysis of the Problem Area}
\label{sec:ConclusionsFromTheAnalysisOfTheProblemArea}

To turn to the original discussion area - privacy in eLearning - it has to be determined that taking care for privacy is a personal concern and should be self-determined by the user him/herself. Thus, the designs of learning environments have to correspond to legal regulations and the users' requirements, accordingly. Consequently, referring to local privacy, every participant of a learning environment should have an own space (cf. the room metaphor) where he/she can adjourn to work on tasks without disturbance by others. In addition, privacy of actions and behaviour can be supported, e.g., by providing sets of different possibilities  for self-determined learning, teaching and creating material. Informational privacy can be ensured by appropriately protecting stored private information and providing the owner of that information the possibilities to regulate its disclosure. A further approach to enable informational privacy consists in offering mechanisms for pseudonymity or anonymity in the learning environment. Teltzrow and Kobsa even suggest a browsing system, which advances current approaches by integrating comprehensive explanations and options for privacy measures \cite{teltzrow.149}. Thereby, they state that privacy is not a global attitude, instead it depends on the actual context, i.e. the user has to rethink and decide for each new domain his privacy settings.
\section{Research Approach and Analysis of Survey Results }
\label{sec:ResearchApproachAndAnalysisOfSurveyResults}

The motivation section has shown that privacy is of particular importance on the Internet and other parts of the digital world. Therefore, this sphere of privacy requires research by examining as well as by inducing studies, which address this problem field.

Our recherché yielded that privacy-related research studies for the field of eLearning exist only of a very poor number. In order to being able to create an objective image of anticipatory results of our own focused approach as well as being able to impartially compare the results with other studies, we introduce few relevant studies that direct on privacy and data protection albeit not necessarily on eLearning. A problem we found confronted with was that the studies addressing this problem area are not up-to-date (some of them were carried out about 8 years ago). However, in order to have a broader focus, we chose studies for our discussions, which seemed to be either good-practice-representative for the European Union or which provided a first impression of a similarly intended mirror image. This decision was further driven by the recognition that other studies that are more state-of-the-art do simply not exist or provide just too little information.

We conducted a study with the special focus on a privacy related examination of eLearning. For that reason, we created an online survey especially intended for analysis of the behaviour and attitudes of European eLearning users regarding informational privacy. Accompanying, we drew comparisons with the results of related studies where it was feasible.

Following, our considerations start with a description of the survey characteristics themselves. Then we present the central questions and the results of the survey as well as an analysing comparison with related research studies. The examination will be completed with a summarised conclusion.

\subsection{Survey Methodology and Sample Characteristics}
\label{sec:SurveyMethodologyAndSampleCharacteristics}

The survey was conducted with focus on eLearning users and their privacy needs and requirements. Therefore, in order to reach the according stakeholders, i.e. European eLearning users, we decided to create an online survey. Moreover, this decision was also driven by the aspect that the target group should be qualified for work with PCs and the Internet, which is a fundamental requirement for eLearning. Information about the survey and the conditions for participation had been published in blogs, discussion forums and web sites. Participation in the survey was anonymous since it is more probable that the respondents would answer more honest and unforced than in an identifiable environment. An online survey that is self-determined, i.e. without interviewer or supervisor, needs to be comprehensible and it has to consist of easily to interpret questions and tasks. In order to test and to configure the questionnaire for release, it had been pre-evaluated. 

Since the survey was open for public access, we are not able to determine a rate of responses. Concerning the survey methodology, the results are rather a cluster sample than a statistical representative one for Europe. That means that there are clusters of participants from specific regions while we could not reach people from other regions who, consequently, could not take part in the survey. Therefore, the results of the survey represent the attitudes only of the respondents and should rather be understood as a tendency for the European eLearning users.

\begin{figure}[htbp]
	\centering
		\includegraphics[width=0.45\textwidth]{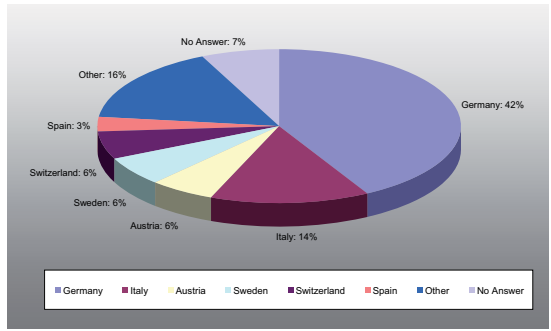}
	\caption{Regional distribution of the survey participants}
	\label{fig:RegionalDistribution}
\end{figure}

The online survey started on 25th November 2005 and ended on 13th January 2006. We got 106 responses from 18 different European countries (cp. Figure \ref{fig:RegionalDistribution}).
 
Regarding the gender distribution of the sample, 66\% of the respondents were male and 28\% were female. The respondents comprised all age groups from 15 to 64 years; but most of them are between 25 and 34 years. The highest completed educational level (we used the graduation model of Great Britain) of the respondents is composed in the following way: the majority of the respondents hold a university degree (32\%) or have already graduated from university (51\%). Thus, most of the respondents are part of the scientific community.

The conducted survey is attended to our central questions divided into five problem fields:
\begin{itemize}
	\item General attitudes,
	\item Dependencies of privacy concerns,
	\item Level of comfort with respect to stating personal information while online,
	\item Demands and provisions of personal information,
	\item Past behaviours and further attitudes.
\end{itemize}

\subsection{General Attitudes}
\label{sec:GeneralAttitudes}

\begin{figure}[htbp]
	\centering
		\includegraphics[width=0.45\textwidth]{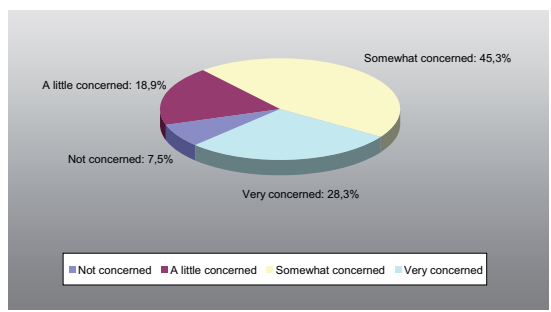}
	\caption{Distribution of levels of concern regarding the protection of privacy data}
	\label{fig:LevelOfConcern}
\end{figure}

One of the first steps of the online survey was to ask about the feelings regarding the security of own personal data while using the Internet (cp. Figure \ref{fig:LevelOfConcern}). About 74\% of the respondents replied that they are somewhat or very concerned. In contrast, only 7.5\% are not concerned.
 
Already, in 2003, a survey addressing general opinions and attitudes of the citizens of the European Union called Eurobarometer 60.0 \cite{soufflotdemagny.159}, considered the field of personal information and privacy in the information society. The Eurobarometer in general represents one of the most important opinion surveys of the European Union. About one thousand inhabitants of each European country take part in these surveys, every time.

The Eurobarometer 60.0 of year 2003 indicated that about 60\% of the respondents were very or fairly concerned of the safety of their privacy when private or public organisations store personal information. 64\% of the respondents "tended to agree that they were worried about leaving personal information on the Internet". The participants were also asked if they should be informed of the reasons why organisations collect personal data and if organisations intend to share such data with other organisations. The vast majority of the respondents (91\%) tended to agree to this question. This statement perfectly reflects the desire for informational self-determination of the citizens of the European Union.

With respect to the socio-demographic analysis, there were little indications that older people and those with a higher level of education were more worried about these issues. But, a distinction between the genders could not be determined.

Misuse of sensitive personal data as well as adaptive advertising on the Internet very much increased in the last years. Accordingly, another relevant study by Cranor et al. \cite{cranor.157} surveyed the concrete needs and raised fears of Internet users, in 1999. Their research focused on acceptance tests and evaluation of the P3P \footnote{http://www.w3.org/P3P/} design where P3P should improve the awareness of the user regarding privacy terms of web sites. Thus, they conducted an online survey with participants of the DRI (Digital Research Inc.) Family Panel. About 500 members from the United States took part in that survey. Cranor et al. draw comparisons to a study of Westin and Harris \cite{harrisassociates.162}, which was published in 1998 with a representative sample of citizens from the USA.
81\% of the respondents of the sample of Westin and Harris and 87\% of the Cranor et al. sample specified that they were very or somewhat concerned about threats to their privacy while online. Cranor et al. explained the distinction of the results by the different compositions of the sample groups. The respondents of the Cranor survey were of higher level of education and possessed more Internet experiences. So, the Cranor results could be understood as an indicator of the "future Internet user population".

The results of the indicated surveys show that there are concerns regarding the safety of personal information and data traces on the Internet. Not only a few people doubt that their data is well protected on the Internet, instead it is the majority. The findings of our study corroborate this conclusion also for the asked eLearning users. 

\subsection{Dependencies}
\label{sec:Dependencies}

The considerations about general attitudes have shown that especially experienced Internet users are most concerned about loosing their privacy. This section will examine if there are special dependencies that influence the level of concern.

The first question we address is: Do privacy interests regarding the online world depend on special knowledge in this field? In order to get an answer on this question, we gave the participants four different terms where three of them were directly related to the field of privacy. The fourth term was an invented acronym and it was used as check term. The respondents of our survey were asked to judge their knowledge of privacy issues by indicating their knowledge of the underlying concepts of the terms. 
The results showed that 40\% of the respondents could explain all three correct terms and 12\% did not pass the check term. It was obvious that the participants with little knowledge of privacy concepts tended to choose the statement "somewhat concerned" (78\%). Whereas the group of respondents with high knowledge in this area tended to the extremes by choosing the statements "not concerned" (10\%) and "very concerned" (31\%).

A further relevant fact of our survey is the examination of dependence of privacy concerns on roles. In the survey, we referred to and explained five different eLearning related roles. The respondents should choose which of the following roles they most frequently hold: 
\begin{itemize}
	\item learner,
	\item tutor (adjustment of teaching material and supervision of  training),
	\item mentor (experienced advisor for individual support),
	\item author (preparation of teaching material for Internet presentation),
	\item administrator (data and user management as well as assignment of user rights).
\end{itemize}
The analysis of the results yielded that 36\% of the participants classified themselves as learners, 15\% as tutors, 5\% as mentors, 26\% as authors, and 6\% as administrators.
The comparison of the according roles with the degree of concern shows that the most concerned group is that of tutors. 87.5\% of them were somewhat or very concerned followed by mentors (80\%) and learners (71.1\%). Administrators showed as few concerns (66.6\%) as authors (67.9\%). 
In fact , in order to allow for, e.g., reasonable adjustment of teaching material by tutors or finding suitable learning groups, information about learners are of special interest. However, their level of solicitude with respect to weakening their privacy is rather low. The reason for this may be due to past grinded in habits, i.e. they most probably expect that other people need their data without thinking about the aims of data gathering. An aspect, which could not be examined in the scope of this study but which might have to be considered, consists in the possible dependence of concern for privacy on the actual learning context: i.e. learning for the job or for leisure. According to a study by Kobsa \cite{patil.163}, the perception of privacy concerns can vary, i.e. in job related terms (like further education or an employment as a tutor), the level of privacy concerns is lower than in leisure related terms.
Furthermore, during analysis of this study, we found interesting that only 14 learners and two authors specified their role as the only role they ever hold. All others sometimes or frequently are consigned to other tasks or roles, i.e., they mostly do not possess a certain role but often hold nearly all of them. Obviously, this diffuse role structure makes it difficult to clearly reason how the dependence of the concerns regarding privacy on the Internet on role assignments looks like.

The dependency of origin of the respondents and their level of solicitude regarding informational privacy is a further interesting point of our survey analysis. Admittedly, even if it is not possible to draw conclusions about single countries on basis of the results of our own survey, it is possible by regions of Europe. That way the results turned out to be as follows: the concerns regarding privacy increased from Southern and Eastern Europe (about 50\% to 64\% were very or somewhat concerned) to Northern and Western Europe (where about 80\% of the according participants were very or somewhat concerned). Thereby, an interesting point is the fact that the same characteristics can be found in the concentration of Internet connectivity of households \cite{schafer.158}. Thus, in 2005, the Eastern and Southern European countries had the lowest Internet concentration (about 20\% to 30\%). In contrast, the Northern and Western European countries had the highest Internet concentration (about 65\% to 85\%), whereby the average rate of all the 25 European countries was 48\%. Hence, we noticed a direct proportionality between the privacy concerns and the Internet concentration: an increased Internet concentration is directly connected to a higher level of concerning with respect to informational privacy regarding the survey results.

\subsection{Convenience Stating Personal Information}
\label{sec:ConvenienceStatingPersonalInformation}

Typically, many different types of personal information are transferred via the Internet. In a next step we want to examine the sensibility of personal information, i.e. we want to clarify which kind of personal data requires more protection in comparison to another.

\begin{figure}[htbp]
	\centering
		\includegraphics[width=0.45\textwidth]{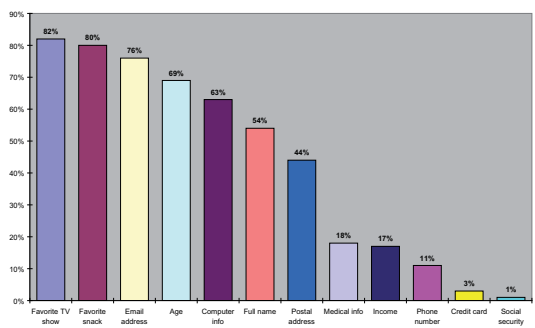}
	\caption{Proportion of respondents (in percent) who feel always or usually comfortable providing specific personal information \cite{cranor.157}}
	\label{fig:ComfortLevelData}
\end{figure}

As Cranor et al. already proved in \cite{cranor.157}, the level of comfort indicating personal information on web sites across various types of data varies very much (cf. Figure \ref{fig:ComfortLevelData}): Accordingly, providing information about preferences, such as favourite TV show or snacks, is always or usually comfortable for the majority (82\% and 80\%) of the interviewees. Similarly, many persons feel comfortable when providing their e-mail address (76\%), their age (69\%), and information about their computers (63\%). In contrast, revealing direct identifying information, i.e. full name and postal address, is for half of the respondents not at ease. In large part, the respondents worried to provide medical information and information about their income (only 17\%--18\% felt always or usually comfortable to provide such information). Only 11\% of the respondents were always or usually willing to provide their telephone number on web sites and merely 3\% would provide their credit card number without doubts. 
Regarding the type of information which serves as contact possibility (telephone, e-mail address and postal address), there are big differences in the level of comfortableness when the persons are asked to indicate them. For the vast majority, it poses no problem to provide the e-mail address (76\%), but supplying the telephone number is comfortable for only 11\% of the respondents. Already in 1991, Westin \cite{westin.161} discovered this issue: He concluded that the level of inconvenience increases the less possibilities an individual has to avoid an undesired communication.

The results of our survey mirror showed an equal situation: credit card number, telephone number, income and postal address are uncomfortable to reveal on the Internet, whereas statements about the age are most comfortable. The indication of school marks is almost as uncomfortable as the postal address, but showing information about school graduation is considered less uncomfortable. 

\begin{table*}[htbp]
	\centering
		\begin{tabular}{|l||c|c|}\hline
			\multicolumn{1}{|c||}{\textbf{Information}}&\multicolumn{1}{c}{\textbf{Internet (average rating)}}&\multicolumn{1}{|c|}{\textbf{eLearning (average rating)}}{\rule[-3mm]{0mm}{8mm}}\\
			\hline\hline
			Credit card number & 6,66 & \\
			Private telephone number & 6,48 & 5,99\\
			Income & 6,31	& \\
			Postal address & 5,98 & 5,59\\
			School marks & 5,48 & 5,19\\
			Personal e-mail address & 5,07 & 3,46\\
			Favourite Internet store & 4,88	& \\
			Occupation & 4,52 & 3,99\\
			Hobbies & 4,37 & 4,3\\
			Information about your PC & 4,18 & 3,69\\
			School qualification & 4,06 & 3,56\\
			Age & 3,29 & 3,01\\
			\hline
		\end{tabular}
	\caption{Average rating of various personal information in a scale of 1 (very comfortable) to 7 (very uncomfortable)}
	\label{tab:AverageRatingOfVariousPersonalInformationInAScaleOf1VeryComfortableTo7VeryUncomfortable}
\end{table*}
	
		These results, which represent a mirror of the users' behaviour w.r.t. indicating different personal information on the Internet, were used to compare with ratings applied especially to the eLearning domain. The results are indicated in Table \ref{tab:AverageRatingOfVariousPersonalInformationInAScaleOf1VeryComfortableTo7VeryUncomfortable}. Hence, one can recognise that the respondents more often would reveal personal data within the eLearning context than in the Internet context, albeit with only slight deviations. However, there is one exception regarding the answers given on "Personal e-mail address": There appeared a rather large distinction between the Internet and eLearning contexts. In addition, the respondents regard revealing the e-mail address as more comfortable than the postal address. Additionally, the indication of telephone numbers is regarded as especially uncomfortable.

\subsection{Demands and Provisions of Personal Information}
\label{sec:DemandsAndProvisionsOfPersonalInformation}

After learning about the users' attitudes regarding the indication of different kinds of personal information in the online world, we like to set those findings in contrast to the question what kind of information is really needed in an eLearning environment. For that reason, the survey participants were asked to decide what kinds of information about other participants they would need for an optimal work depending on the own role as well as on the other roles. Moreover, the participants should also decide which of the following kinds of personal information they think that other participants would need for an optimal work: 
\begin{itemize}
	\item e-mail address or telephone number,
	\item interests,
	\item postal address and date of birth,
	\item competences \& completed educations,
	\item automatically recorded activities within the eLearning application,
	\item name and picture of the person as well as
	\item information especially related to learners:
	\begin{itemize}
		\item learning aims \& motivation,
		\item preferred presentation of teaching material, and
		\item achieved scores in eLearning tests.
	\end{itemize}
\end{itemize}

The responses indicated that, in order to establish an optimal work situation, most of the participants would provide more personal information as they would demand from others. The reason for this interesting result may be seen in the experiences the participants already had attained during their work with eLearning environments: as we will see below, efficient collaboration and cooperation are only possible if the users are aware of each other. 

Out of the above-mentioned items, the participants indicated learning targets, competencies and contact information as most important information about a learner. The most relevant facts about the other roles are a visual impression of the respective person, information about how to contact this person, and his/her competencies. However, there are also discrepancies: According to the results of the survey, learners provide their learning targets more likely to tutors than to other learners or authors. That behaviour may originate in a possible need for adjustments of learning material or the kind of mentoring by the tutor, respectively.
As mentioned above, the majority of the respondents do not take only one single role in an eLearning environment. Instead, sometimes they possess even all of the roles. So, the mutual trust in these roles is afflicted. A potential solution to circumvent this problem could be the usage of different pseudonyms/identities when changing a role.

As already mentioned, in order to allow for cooperative teamwork, a certain degree of group awareness is necessary. Users need to know that other people are available for interactions and they should know who are those people. A study, which was concerned with analysing users' privacy attitudes in eLearning [Kettel et al.], started with recognising that a large number of users set themselves invisible for the virtual community. In the beginning of the study, those users were not willing to reveal any personal information to other users. However, they were very interested in gathering information about the others. In order to encourage the users to disclose personal information, the survey organisers set up the rule that each user gets to see only those kinds of information about other participants that they reveal about themselves to the others. After this rule was established, the users disclosed more information about themselves and, thus, they fostered the formation of the learning community.
After having an insight into the online activities of the other participants, the participants were asked if they assess it advantageous being able to watch those activities. 53\% of the participants responded with "yes" while 12.5\% of the participants saw no benefit. Further, 87.5\% agreed that the potential advantages of sharing personal information outweigh the potential risks for privacy. Therefore, the vast majority of the participants would make a trade-off in favour of the awareness of the community.

Looking at the results of our own study and the study of Kettel et al., we can conclude that information about other participants of an eLearning environment could improve the behaviour of the whole group. However, it also influences the informational privacy of the individual.

\subsection{Past Behaviour and Further Attitudes}
\label{sec:PastBehaviourAndFurtherAttitudes}

In the last block of survey questions, we approached the research question if respondents' behaviour on the Internet matches with their privacy concerns. The majority of the respondents of our survey indicated that they were very or somewhat worried about the safety of their informational privacy. However, do they also try to protect their personal data stored on their computer?

The participants were asked if the computer they use for working with the Internet has firewall and virus scanner software installed. As expected, the respondents who are not or less concerned do less provide for a risk (87.5\%) than very concerned respondents (100\% of those participants have installed a firewall and 93.4\% of them - a virus scanner). Nevertheless, the majority of the respondents generally use such kind of software. 
Another aspect was to learn about the participants' practice of regularly changing important passwords. Thus, we asked the respondents if they had changed their important passwords within the past six months. Again, we recognised that with an increase of the users' concerns also the frequency of password changes increased. Thus, only 37.5\% of the respondents who are not worried indicated to do it. In contrast, 53.4\% of the very concerned respondents pointed out to care for regular password changes. However, the relatively small amount of only half of the very concerned respondents could be a result of the difficulties connected with memorising safe passwords for various applications. 
Another question addressed the users' attitude of making wrong statements in online registrations to protect their privacy. 80\% of the participants responded to this question with "yes". Additionally, there was a small increase from the proportion of the not concerned respondents (75\%) to one of the somewhat and very concerned respondents (about 82\%). 
Further, the respondents ranked the trust in the provider of a website among the most important facts when they are asked to fill in an online registration. Following, the participants indicated the worthiness of the registration's benefits as well as the indication of a reasonable purpose of gathering the personal data.
A further relevant aspect is the use of different e-mail accounts. At this point, we noticed a huge variance of behaviour between the differently concerned people. Thus, while 93.3\% of the very concerned respondents use different e-mail accounts for several spheres of life, 62.5\% of the not concerned participants do not separate their e-mail communications by different accounts.

Turning back to the specifics of user behaviour in eLearning applications, we presented the participants a set of different scenarios that might relate to privacy concerns in an eLearning environment. The survey participants should decide if they either favour or oppose the following functions of an eLearning environment:
\begin{enumerate}
	\item Automatic notice of false test results to the tutor;
	\item Possibility to learn anonymously in an eLearning environment;
	\item Creating several accounts and nicknames for different courses; 
	\item Logging on the computer does log the user in to further applications like e-mail account and eLearning  application, at the same time;
	\item Getting access to a course, which has proper examples adapted to the user's education, job and previous knowledge, in return for the indication of personal data;
	\item Function to display which of the private and identifying data is disclosed to different groups of users and which of these data is hidden from them;
	\item Private and identifying data is handed to other users only with the explicit permission of the owner;
	\item All private and personal data can be seen by everyone being part of the eLearning application.
\end{enumerate}

\begin{figure}[htbp]
	\centering
		\includegraphics[width=0.45\textwidth]{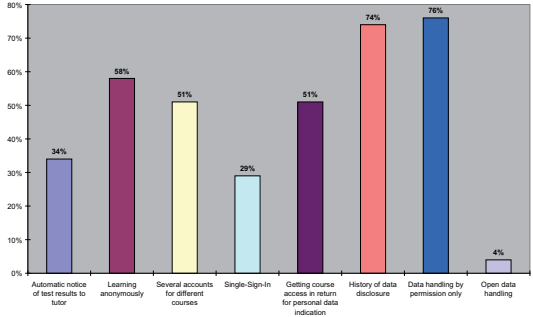}
	\caption{Proportion of respondents (in percent) who favoured specific features}
	\label{fig:Features}
\end{figure}

As shown in Figure \ref{fig:Features} the functions (2) and (3) display anonymous and pseudonymous learning and are favoured by about half of the respondents. In contrast, about 50\% of the participants would approve personalisation (5). Only few of the respondents endorse the functions (1) (34\% of the respondents) and (4) (29\%). The most important functions (6) and (7), which enable the learner to regulate his/her private information and thus to establish informational self-determination, are favoured by about 75\% of the respondents. Consequently, more than 95\% of the participants disliked a form of data handling where everyone would be able to access all information.

\subsection{Summary}
\label{sec:Summary}

In result of the survey analysis, we record that the sample of our study shows a high level of concern regarding informational privacy among eLearning users. Especially, the respondents who have good knowledge in privacy protection more often tend to the extremes of solicitude than others. Based on our results, evidence suggests that an increase of concern related to privacy can be noticed depending on the regionality the respondents resided in. Hence, respondents from Eastern and Southern Europe are less often worried of their privacy in comparison to the participants from Western, Middle, and Northern Europe. In view of eLearning roles, tutors and mentors are the groups with the highest proportion of concerned respondents followed by the learners. Personal information is differently handled when revealed to others. Unproblematic data is, e.g., the age followed by the e-mail address. Most of the respondents avoid specifying their private telephone number in online communications. In an eLearning environment, the respondents indicated that they would provide more personal information about themselves than they would ask from other participants. With respect to the past behaviour, respondents who showed a high level of concern more often use possibilities to protect their personal data than users with fewer concerns. According to the analysis of the attitudes towards possible privacy protecting measures, 95\% of the respondents would particularly prefer control mechanisms, which allow access to personal data by others only with the explicit permission of the respective owner.
As already mentioned, the study does not represent a statistically representative sample for Europe since the analyses of demographic aspects of the responses showed that the majority of the survey's participants is working in an academic environment. In addition, no equal distribution regarding the origin countries could be achieved. Therefore, the results of the survey represent the respondent's attitudes and should be understood as a tendency for the European eLearning users. 

\section{Conclusion}
\label{sec:Conclusion}

Most of the presented studies examine privacy within the context of the Internet. By comparing these studies with our own survey results, we were able to analyse if there are differences between the group of Internet users and the group of eLearning users regarding their privacy needs. Thus, we could confirm that the majority of the respondents, which are eLearning users, see a need in protecting their informational privacy also within an eLearning environment. However, in order to reasonably interact with each other, eLearning users are also aware of the necessity to disclose parts of their personal information.

Based on the results of the survey analysis, we expect that future interests head for adaptive support in order to optimise eLearning offers. Therefore, several personal data will be required and the eLearning users - especially the learners - will have to reveal data about themselves. In addition, activity data while using the eLearning offer is and will be stored, too. This trend will make it more difficult to protect private information. Users who care for their privacy must have the possibility to come to know of this as well as to determine by themselves how their personal trade-off between privacy and utilisation of services should look like.
With the increase of the number of people using eLearning, the influence of social factors will increase, as well. Thus, a community, the members of which want to learn together, will need mechanisms of competence rating, in order to being able to assess the interaction partners. In addition, communication, cooperation, and collaboration have to be easily usable. This also includes enhancements regarding the offered functionality of group awareness by respecting privacy issues. 

By researching needs and requirements regarding privacy, we have come to know, that privacy and informational self-determination are required or even demanded, respectively, by the majority of eLearning users. A next step will be to examine present eLearning environments with regard security and privacy regulations. Furthermore, our research team is currently developing an eLearning environment called BluES'n \cite{borceapfitzmann.31}. The further realisation of that platform is being carried out following user-determined privacy regulations by respecting and considering the results gained in the survey documented in this paper.

\section{Acknowledgements}
\label{sec:Acknowledgements}

We like to thank Elke Franz and Katja Liesebach for helpful discussions, and all in the survey involved people for their support and encouragement. Special thanks go to the participants of the survey who, in the first place, made our work possible by responding.
The work reported in this paper was supported by the IST project PRIME; however, this work not necessarily represents the view of the project.

\bibliographystyle{alpha}
\bibliography{TR_Privacy_in_eLearning}

\begin{thebibliography}{BPLP05}

\bibitem[BPLP05]{borceapfitzmann.31}
Katrin Borcea-Pfitzmann, Katja Liesebach, and Andreas Pfitzmann.
\newblock Establishing a privacy-aware collaborative elearning environment.
\newblock In {\em EADTU Working Conference 2005}, Rome, 2005. EADTU.

\bibitem[CRA99]{cranor.157}
Lorrie~Faith Cranor, Joseph Reagle, and Mark~S. Ackerman.
\newblock Beyond concern: Understanding net users' attitudes about online
  privacy, 1999.

\bibitem[EPC95]{anon.160}
{Directive 95/46/EC of the European Parliament and of the Council: On the
  protection of individuals with regard to the processing of personal data and
  on the free movement of such data}.
\newblock Legal Ruling/Regulation, 1995.

\bibitem[H{\&}AW91]{westin.161}
Louis Harris {\&}~Associates and Alan~F. Westin.
\newblock Harris-equifax consumer privacy survey, 1991.

\bibitem[H{\&}AW98]{harrisassociates.162}
Louis Harris {\&}~Associates and Alan~F. Westin.
\newblock E-commerce and privacy: What net users want.
\newblock Technical report, Price Waterhouse and Privacy {\&} American
  Business, 1998.

\bibitem[KBG04]{kettel.94}
Lori Kettel, Christopher Brooks, and Jim Greer.
\newblock Supporting privacy in e-learning with semantic streams.
\newblock In {\em Second Annual Conference on Privacy, Security and Trust},
  pages 59--67, New Brunswick, Canada, 2004. University of New Brunswick.
\newblock Several surveys to determine the willingness of learners in sharing
  their online activities.

\bibitem[OEC03]{anon.164}
{OECD Guidelines: On the Protection of Privacy and Transborder Flows of
  Personal Data}.
\newblock Government Report, 2003.

\bibitem[PH06]{pfitzmann.117}
Andreas Pfitzmann and Marit Hansen.
\newblock Anonymity, unlinkability, unobservability, pseudonymity, and identity
  management: A consolidated proposal for terminology, 2006.

\bibitem[PK04]{patil.163}
Sameer Patil and Alfred Kobsa.
\newblock Instant messaging and privacy.
\newblock In {\em Human Computer Interaction 2004}, pages 85--88, Leeds, UK,
  2004.

\bibitem[R{\"o}s01]{rossler.165}
Beate R{\"o}ssler.
\newblock {\em Der Wert des Privaten}.
\newblock Suhrkamp, Frankfurt/Main, 2001.

\bibitem[SCFF05]{schafer.158}
G.~Sch{\"a}fer, S.~Cervellin, M.~Feith, and M.~Fritz, editors.
\newblock {\em Europe in figures: Eurostat yearbook 2005}.
\newblock Eurostat yearbook. Office for Official Publications of the European
  Communities, Luxembourg, 2005.
\newblock P. 220.

\bibitem[SdM03]{soufflotdemagny.159}
Renaud Soufflot~de Magny.
\newblock Eurobarometer 60.0: Consumer rights, data protection, education
  through sport, product safety, e-commerce, attitudes towards people with
  disabilities, and the euro.
\newblock Government Report, 2003.

\bibitem[TK04]{teltzrow.149}
Maximilian Teltzrow and Alfred Kobsa.
\newblock Communication of privacy and personalization in e-business.
\newblock In {\em 1st Workshop {"}A Multiple View of Individual Privacy in a
  Networked World{"}}, January30--31 2004.

\bibitem[VZU83]{anon.1}
Volksz{\"a}hlungsurteil.
\newblock Legal Case, 1983.

\end{thebibliography}

\end{document}